# Title: Self-driving laboratory for accelerated discovery of thin-film materials

**One sentence summary:** The first autonomous laboratory for the discovery of thin films is used to optimize the doping and annealing of organic semiconductors.

**Authors:** B. P. MacLeod[1,3†], F. G. L. Parlane[1,3†], T. D. Morrissey[1,3], F. Häse[4-7], L. M. Roch[4-7], K. E. Dettelbach[1], R. Moreira[1], L. P. E. Yunker[1], M. B. Rooney[1], J. R. Deeth[1], V. Lai[1], G. J. Ng[1], H. Situ[1], R. H. Zhang[1], M. S. Elliott[1], T. H. Haley[1], D. J. Dvorak[3], A. Aspuru-Guzik[4-10]*, J. E. Hein[1]*, C. P. Berlinguette[1-3,8]*

**Affiliations:**

[1]Department of Chemistry, The University of British Columbia, Vancouver, British Columbia, Canada
[2]Department of Chemical & Biological Engineering, The University of British Columbia, Vancouver, British Columbia, Canada
[3]Stewart Blusson Quantum Matter Institute, The University of British Columbia, Vancouver, British Columbia, Canada
[4]Department of Chemistry and Chemical Biology, Harvard University, Cambridge, Massachusetts, USA
[5]Department of Chemistry, University of Toronto, Toronto, Ontario, Canada
[6]Department of Computer Science, University of Toronto, Toronto, Ontario, Canada
[7]Vector Institute for Artificial Intelligence, MaRS Centre, Toronto, Ontario, Canada
[8]Canadian Institute for Advanced Research (CIFAR), MaRS Centre, Toronto, Ontario, Canada
[9]Kebotix, Inc., 501 Massachusetts Ave., Cambridge, Massachusetts, USA
[10]Zapata Computing, Inc., 100 Federal St, 20th Floor, Boston, Massachusetts, USA

[†]These authors contributed equally to this work

*Email: cberling@chem.ubc.ca, alan@aspuru.com, jhein@chem.ubc.ca

**Abstract:** Discovering and optimizing commercially viable materials for clean energy applications typically takes over a decade. Self-driving laboratories that iteratively design, execute, and learn from material science experiments in a fully autonomous loop present an opportunity to accelerate this research. We report here a modular robotic platform driven by a model-based optimization algorithm capable of autonomously optimizing the optical and electronic properties of thin-film materials by modifying the film composition and processing conditions. We demonstrate this platform by using it to maximize the hole mobility of organic hole transport materials commonly used in perovskite solar cells and consumer electronics. This demonstration highlights the possibilities of using autonomous laboratories to discover organic and inorganic materials relevant to materials sciences and clean energy technologies.

**Main text:**

**Introduction**

Optimizing the properties of thin films is time intensive because of the large number of compositional, deposition, and processing parameters available (*1*, *2*). These parameters are often correlated and can have a profound effect on the structure and physical properties of the film and any adjacent layers present in a device (*3*). There exist few computational tools for predicting the properties of materials with compositional and structural disorder, and thus the materials discovery process still relies heavily on empirical data. High-throughput experimentation (HTE) is an established method for sampling a large parameter space (*4*, *5*), but it is nearly impossible to sample the full set of combinatorial parameters available for thin films. Parallelized methodologies are also constrained by the experimental techniques that can be used effectively in practice. The overwhelming size of the thin-film materials parameter space motivates the need for both data- and theory-guided algorithms for executing experiments beyond what can be achieved with HTE alone (*6–8*).

The experimental approach of iterating between automated experimentation and machine-learning-based experiment planning has resulted in early successes in addressing high-dimensional problems in experimental physics (*9*), chemistry (*10*), and life-sciences (*11*). This approach is only starting to be implemented in the materials sciences (*1*), as demonstrated by the optimization of carbon nanotube growth (*12*), amorphous alloy compositions (*13*), and inorganic perovskite quantum dot nucleation (*7*). We demonstrate here the optimization of thin films using our platform named *"Ada"*, a flexible and modular self-driving laboratory capable of autonomously synthesizing, processing, and characterizing organic thin films (Fig. 1; Fig. S1; Movie S1). *Ada* trains itself how to find target parameters without any prior knowledge, enabling iterative experimental designs that maximize the information gain per sample (Fig. 2).

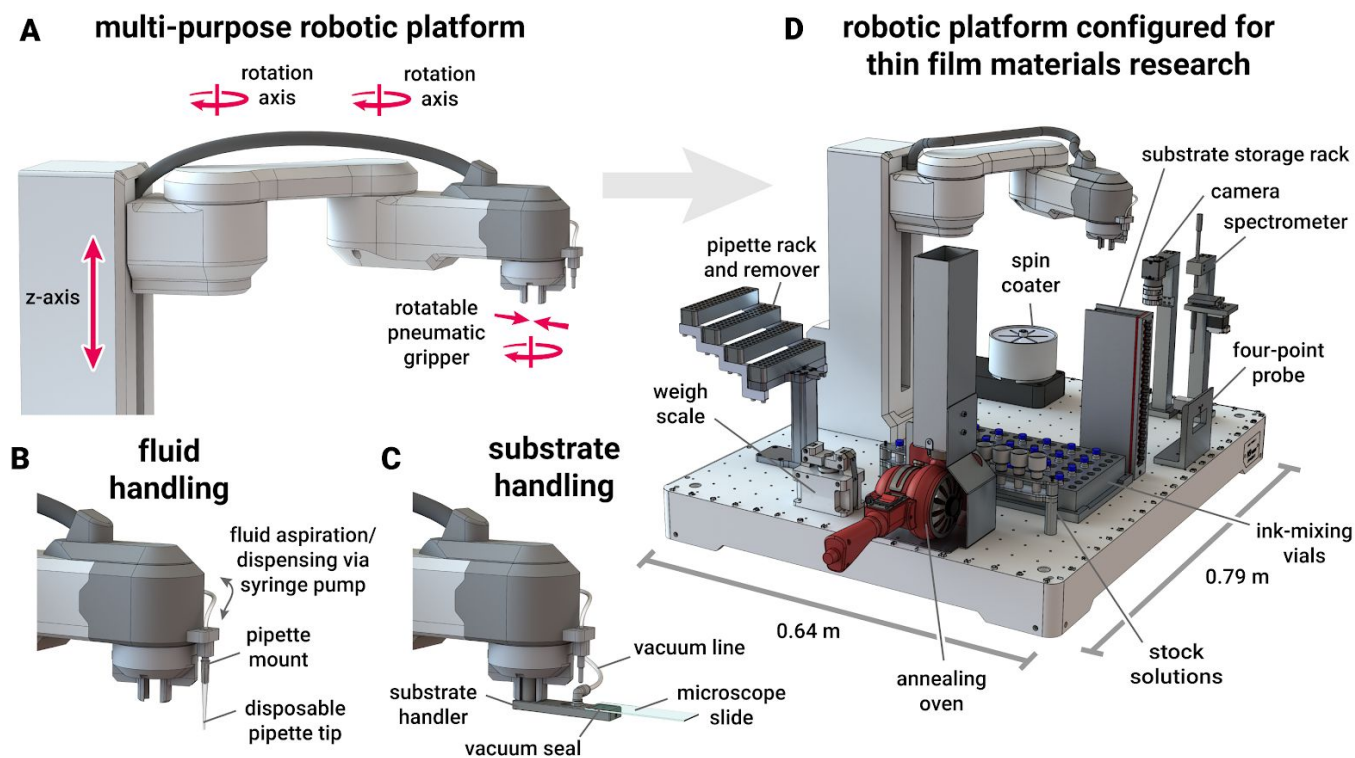

**Fig. 1. The *Ada* self-driving laboratory.** (**A**) The self-driving laboratory is based on a modular robotic platform which interacts with objects using a rotatable pneumatic gripper on a polar robotic arm achieving 10 μm repeatability and ~1 m/s maximum velocity. (**B**) Fluid handling is achieved using disposable pipette tips which can be press-fit onto and removed from the arm's pipette mount by the robot. Pipetting with a mean accuracy of 5 μL is achieved using a syringe pump connected to the pipette mount. (**C**) Substrate handling is achieved using a vacuum substrate handler gripped by the robotic arm. (**D**) Configuration of the robotic platform for a specific experimental workflow is achieved by mounting an appropriate collection of experimental modules on the robot; here the *Ada* platform is shown equipped for the synthesis and characterization of thin film materials.

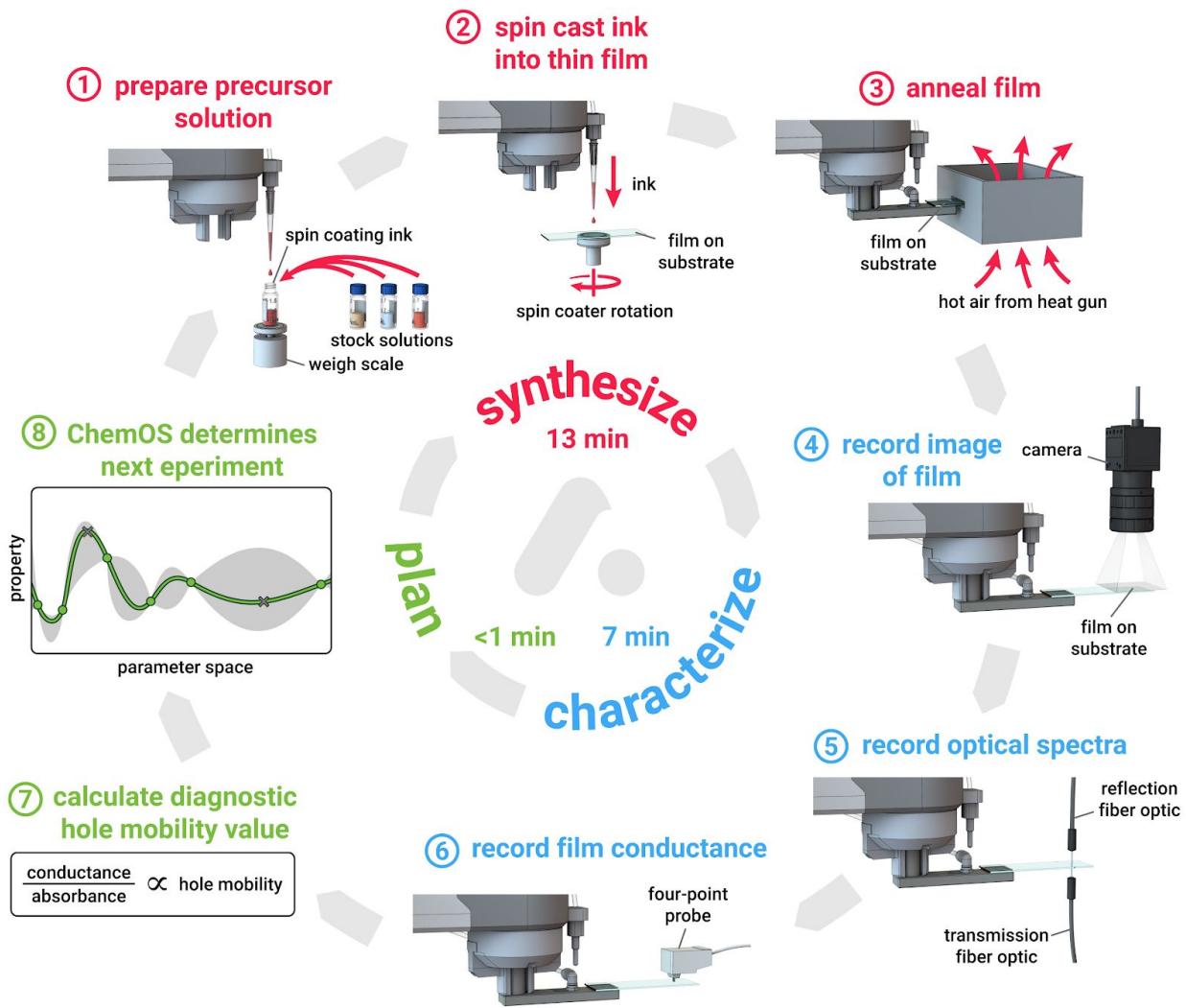

**Fig. 2. *Ada* employs an autonomous optimization workflow.** The autonomous workflow involves iterative experimentation with the goal of discovering a thin-film composition with the highest possible "pseudomobility". Each iteration of the workflow involves: [1] mixing an HTM-dopant-additive ink, [2] spin coating the ink onto a substrate, [3] thermally annealing for a variable amount of time, [4] imaging with a visible-light camera, [5] acquiring UV-Vis-NIR spectra in reflection and transmission modes, [6] measuring the I-V curve of the film with a 4-point probe, [7] computing a pseudomobility based on the IV and spectroscopic data, and [8] feeding this pseudomobility into the *ChemOS* (*14*) orchestration software and the *Phoenics* (*15*) Bayesian optimization algorithm which then designs the next experiment.

**Results**

As a first step in proving out the methodology, we designed *Ada* to target organic hole and electron transport layers that are ubiquitous in advanced solar cells (*16*), as well as optoelectronics applications such as organic lasers (*17*) and light emitting diodes (*18*). For this work, we configured *Ada* specifically to optimize the hole mobility of spiro-OMeTAD, an organic HTM common to perovskite solar cells (PSCs) (*19*). The hole mobility of spiro-OMeTAD is critical to PSC performance, but it is highly sensitive to dopants, additives, spin-coating solvents, and post-deposition processing (*19–26*). How each of these factors affect the hole mobility of amorphous spiro-OMeTAD remains difficult to model (*3*, *27*), and thus optimizing the relevant properties of spiro-OMeTAD is still done empirically. This optimization process often takes months to complete and slows the translation of new organic hole and electron transport layers for solar cells and related devices.

*Ada* autonomously optimizes the hole mobility of spiro-OMeTAD by: (i) measuring and mixing solutions of HTMs, dopants, and plasticizers; (ii) depositing solutions as thin films on rigid substrates; (iii) annealing each film for a specified duration (iv) imaging each film to detect morphologies, defects, and impurities; and (v) characterizing the optical and conductivity properties to produce surrogate hole mobility data. This data is received by *ChemOS* (*14*), which uses the *Phoenics* (*15*) global Bayesian optimization algorithm to design new experiments by actively learning from previously acquired data. *Phoenics* uses a sampling parameter to explicitly bias experimental design towards exploration or exploitation in an alternating fashion, and has been shown to outperform random and systematic searches (*14*, *15*, *28*). The optimization experiments are performed by a multi-purpose robot (Fig. 1) equipped with a rotatable pneumatic gripper and a pipette mount which enable the platform to accomplish a wide variety of tasks by interacting with a number of different modules. The platform aspirates, dispenses, and mixes liquid precursors with the assistance of syringe pump and a weigh scale. Precursor solutions are spin-cast as thin films on glass substrates, which can then be annealed up to 165 °C (Fig. S2) using a forced convection annealing system which enables control over the extent of annealing by leveraging the ability of the robot to accurately and repeatably position the sample in a hot air stream for a precisely controlled duration . *Ada* then characterizes the films using purpose-built systems for dark field photography, UV-Vis-NIR reflection and transmission spectroscopy, and 4-point probe conductance. The robot also serves as a XYZ sample-positioning stage enabling all characterizations to be performed at multiple positions on the sample, which we leverage to collect spectroscopy and conductance data at 7 spatial positions on each sample. One sample is synthesized and characterized approximately every 20 minutes, with consumables (pipettes, substrates, stock solutions) requiring replenishment every 7 samples. The ability to produce high quality, well-organized datasets (Figs. S3 and S4) while also enabling typically uncontrolled variables (e.g., time between process steps, height of spin coating dispense nozzle) to become controlled or optimization parameters are very powerful features of *Ada*. Moreover, *Ada* is controlled using flexible, open-source Python software (see Materials and Methods), which facilitates the rapid implementation of new experiments.

We selected HTM hole mobility as our target parameter for optimization, but this parameter typically requires assembly of multilayer devices in order to get a valid measurement (*27*, *29*, *30*). Conventional methods are simply not compatible with the time scale needed for efficient autonomous optimization (*31*). We therefore developed a scheme where we could use 4-point-probe conductivity and UV-Vis-NIR spectroscopy measurements to produce a diagnostic quantity, "pseudomobility", that is proportional to hole mobility (see Materials and Methods and Supplementary figure S5). Pseudomobility is the quotient of the sheet conductance of a thin film and the absorptance of oxidized spiro-OMeTAD in the film. We estimated the absorptance of each thin film with an analytical model which incorporates experimental reflection and transmission spectra, and which accounts for the effect of the glass substrate on these spectra. The pseudomobility ratio, which provides a thickness-independent low latency analytical surrogate for hole mobility, became our target optimization objective and enabled us to accelerate the rate of meaningful data collection

The pseudomobilities of spiro-OMeTAD thin films were optimized by iteratively designing film compositions with variable annealing times and dopant concentrations (tables S1, S2). Solutions prepared from stock solutions of spiro-OMeTAD and a cobalt(III) dopant (along with a fixed amount of the plasticizer 4-*tert*-butylpyridine) were spin-coated onto substrates to yield thin films. Each film was annealed, imaged, and analyzed to determine a pseudomobility value that was relayed to *ChemOS*. Fig. 3A chronicles how the doping ratios and annealing times were varied during optimization for two independent experimental campaigns. Each of the two 35-sample campaigns took under 30 h (including time for restocking consumables) and each had one failed sample. This failure rate is typical for our system. An important outcome is that both campaigns converged on the same global maximum for both doping ratio (~0.4 eq.) and annealing time (~75 s) to deliver films with the same maximum pseudomobilities. This reproducible endpoint is significant and demonstrates that *Ada* can successfully navigate a broad experimental space (see table S3 for additional reproducibility data).

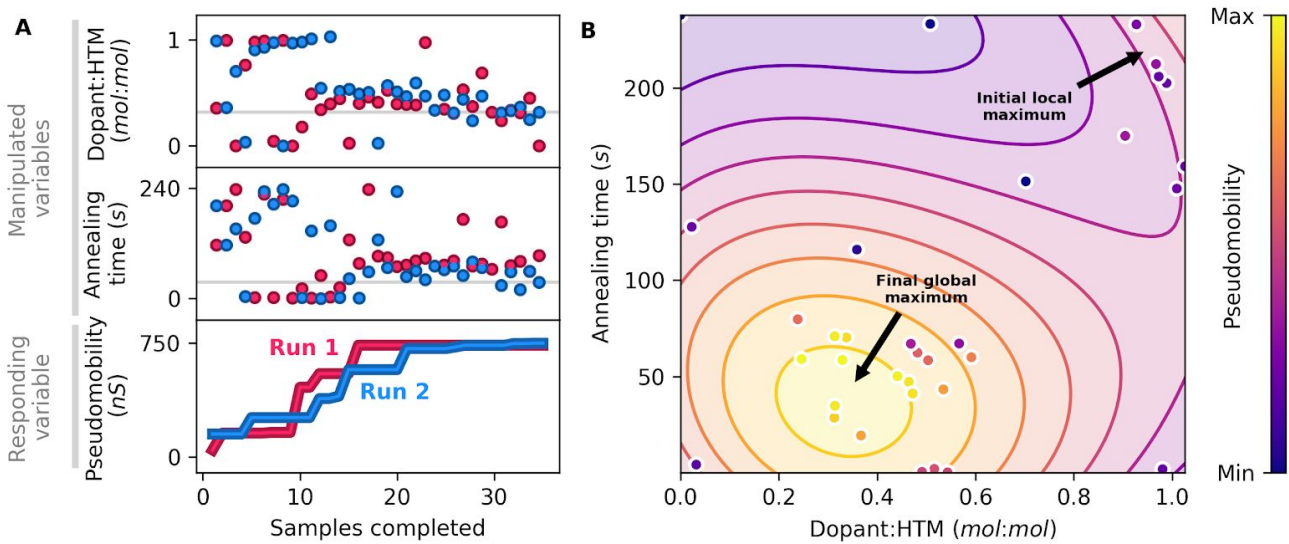

**Fig. 3. Results of thin-film pseudomobility optimization carried out by the self-driving lab.** (**A**) Experimental values for cobalt doping ratio, annealing time, and maximum measured pseudomobility as a function of the number of experiments performed for two independent optimization runs. (**B**) The pseudomobility response surface and sampled points for the second (blue, left) optimization run. The algorithm initially discovered a local maximum, and then discovered the global maximum of the sampled parameter space.

Fig. 3b shows the locations and sequence of the experimentally sampled points in the parameter space. The sampled points can be seen to initially cluster at a local maximum (~100% doping and annealing time >200 s) prior to finding a higher performance region elsewhere in the parameter space. While the eventual rejection of the local maximum confirms that the *explore-exploit* functionality of *Phoenics* can prevent the search from becoming stuck near local optima, we were curious why *Ada* identified a local maximum at high doping levels. Subsequent investigations of the dark field images of these films revealed annealing-induced dewetting of the films containing intermediate amounts of dopant (Fig. S6). At elevated doping levels, dewetting was suppressed, allowing a region of improved thermal stability to be identified (see Supplementary Information). The favourable performance of high dopant/high annealing time films was not intuitive (*32*), and this observation was facilitated by our autonomous platform, which searched over a larger range of doping and annealing conditions than is typically explored in studies of organic HTMs. We hypothesize that the highly doped films are stable at high annealing temperatures because of the greater intrinsic thermal stability of the dopant (which does not readily form a glass and melts at 189 °C, see Fig. S7) compared to amorphous spiro-OMeTAD (which exhibits a glass transition at 124 °C, see (*33*)). On this basis, when sufficient Co(III) salt is added to the HTM film, the increase in the stability of the doped HTM due to the intrinsic stability of the dopant overcomes the decrease in thermal stability generally associated with the addition of dopants to hole transport materials(*32*, *34*, *35*). This result is non-trivial and represents an unexpected scientific observation from an artificially intelligent experimental design.

**Discussion**

We report here the first use of a self-driving laboratory to optimize composition and processing parameters for thin-film materials. This proof-of-principle study targeted the optimization of a type of thin organic semiconducting film common to advanced solar cells, but the modularity of our robotic platform and control software enables the rapid incorporation of new experiments, techniques, analytical hardware, and algorithms. The *Ada* platform can therefore be easily tailored for a range of inorganic and organic materials and applications, and even be coupled to automated organic synthesis methodologies developed for the pharmaceutical industry (*36*, *37*). The next stage of development for this robotic platform is to introduce sequential film deposition to extend autonomous optimization experiments to multi-layered systems that comprise full devices. As the robotic workflow complexity increases, the experimental throughput of our current, fully-serial scheme will decrease; to overcome this challenge, we plan to develop stand-alone synthesis and characterization modules which can run in parallel. Our ability to iteratively modify Ada will prove to be particularly useful in this regard. Indeed, we expect platforms such as *Ada* to facilitate the deployment of effective autonomous experimentation at a scale compatible with the rapidly evolving needs and constraints (e.g., budget, time, space) of a broad cross-section of the materials science research community.

**Materials and Methods**

*Materials*

Acetonitrile (CAS 75-05-8, HPLC grade, ≥ 99.9%), toluene (CAS 108-88-3, ACS grade), acetone (CAS 67-64-1, ACS grade), spiro-OMeTAD (CAS 207739-72-8, HPLC grade, 99%), FK 102 Co(III) TFSI salt (Sigma Aldrich product number 805203, 98%), and 4-tert-butylpyridine (CAS 3978-81-2, 96%) were purchased from Sigma Aldrich and were used without further purification. Extran® 300 detergent (EX0996-1) and 2-propanol (ACS grade, ≥99.5%) were purchased from EMD Millipore Corporation and were used without further purification. Microscope slide substrates (75 × 25 × 1 mm, VWR Cat. No. 16004-430) were purchased from VWR International.

*Consumables for the robotic platform*

During the optimization experiments, the robotic platform was periodically restocked with consumables. These include 3" × 1" × 1 mm microscope slides (VWR VistaVision), 2mL HPLC vials (Canadian Life Science), 200μL pipettes (Biotix M-0200-BC) and various stock solutions. Cleaning of the microscope slides and preparation of the stock solutions are further detailed below.

*Manual preparation of stock solutions*

All reagent solutions were prepared in an atmosphere dried over anhydrous calcium sulfate (DRIERITE®), resulting in ~0.005 mg/L of water remaining in the atmosphere. Toluene and acetonitrile solvents were prepared by drying over anhydrous magnesium sulfate, filtering through a 0.2 μm PTFE filter, and stored over 3 Å molecular seives. A solution of 1:1 v/v solution of acetonitrile/toluene (MeCN/MePh) was prepared by mixing equal volumes of dry acetonitrile and toluene. A stock solution of spiro-OMeTAD was prepared by briefly (1-5 min) sonicating a mixture of spiro-OMeTAD (off-white powder) with MeCN/MePh. The resulting solution had a transparent, pale yellow color. A stock solution of FK 102 Co(III) TFSI salt was prepared by dissolving FK102 Co(III) TFSI salt (bright orange crystalline powder) in MeCN/MePh and stored without exposure to UV-light. The resulting solution had a transparent bright orange color. A stock solution of 4-tert-butylpyridine was prepared by dissolving tert-butylpyridine (clear, colorless solution) in MeCN/MePh. All stock solutions were prepared at concentrations of 50 mg solute per 1 mL solvent.

*Manual cleaning of substrates*

75 × 25 × 1 mm microscope slide substrates were cleaned through multi-step sonication. First, the slides were sonicated for 10 min in a 5% v/v solution of Extran® in deionized water. The sides were then sonicated sequentially in deionized water, acetone, and 2-propanol for 10 minutes each step. The slides were stored in 2-propanol and dried with filtered air before use.

*Robotic methods*

The robot used is a Selective Compliance Assembly Robot Arm (SCARA)-type robot (N9, North Robotics; www.northrobotics.com) which performs the robotic manipulations in our workflows. This robot is driven by a controller (C9, North Robotics), which also provides auxiliary controls for third-party instruments and components used by the robot. The controller and additional peripherals are controlled by a computer running a Python script based on open-source libraries (see https://gitlab.com/ada-chem).

*Autonomous workflow step 1: Robotic preparation of spin-coating inks*

The precursor solution for each sample was prepared by mixing varying amounts of: spiro-OMeTAD stock solution; FK 102 Co(III) TFSI salt stock solution; and 4-tert-butylpyridine stock solution in ambient conditions. In each precursor solution the FK 102 Co(III) TFSI:spiro-OMeTAD ratio (n/n) was between 0 to 1, with the ratio determined by the ChemOS orchestration software. The ratio (m/m) of tert-butylpyridine to the total amount of spiro-OMeTAD and FK102 Co(III) TFSI was fixed at 0.2. The resulting precursor solutions became dark purple in appearance upon the combination of the spiro-OMeTAD and FK 102 Co(III) TFSI solutions. The precursor solutions were mixed through aspiration and were used within a minute of preparation.

*Autonomous workflow step 2: Robotic spin coating of thin-film samples*

The thin-film samples were prepared via spin-coating with a custom-built spin-coater provided by North Robotics. The microscope slides were spun at 1000 rpm and 0.100 mL of the precursor solution was dispensed at a normal incidence at the center of the slide. Rotation continued for 60 s.

*Autonomous workflow step 3: Robotic thermal processing of thin-film samples*

The forced convection annealing furnace was constructed from a MHT Products Inc. model 750 heat gun facing upward into a vertically oriented 75 × 50 mm rectangular aluminum tube kept under ambient conditions. A 40 × 5 mm sample port was cut 40 mm from the heat gun. Freshly spin coated thin-film samples were moved by the N9 slide gripper into the furnace via the sample port, after which the heat gun power was triggered for the amount of time requested by the orchestration software. The temperature profile of the annealing procedure is shown in Supplementary Fig 2. The temperature of the slide ramps from ambient temperature to 165 °C over the first 100 s and remains at that temperature for the rest of the annealing time. After the requested heating time has elapsed, the arm immediately removed the sample from the furnace and held 25 mm above a 4500 rpm cooling fan for 3 min. This cooling period allowed samples to return to ambient temperature, regardless of annealing time, before further characterization.

*Autonomous workflow step 4: Robotic dark field photography*

Thin-film samples were imaged at a dark field photography station composed of a FLIR Blackfly S Mono 12 MP USB Vision (Sony IMX226) camera mounted above an AmScope MIC-209 3 W ring light. The sample was moved by the robotic arm to 90 mm below the camera and illuminated by the ring light to provide contrast between smooth and rough regions of the film. Images were captured at three different overlapping locations at a resolution of 4000 × 3000 px. Manual post-experiment analysis of collected images was used to identify dust, defects, and dewetting in thin-film samples.

*Autonomous workflow step 5: Robotic UV-Vis-NIR spectroscopy*

**Spectrometer design**

UV-Vis-NIR transmission and reflection spectra were collected with a custom-built, fiber-optic spectroscopy station. A BLACK-Comet UV-Vis Spectrometer (190 - 900 nm, < 1 nm resolving resolution), a DWARF-Star Miniature NIR Spectrometer (900 - 1700 nm, 2.5 nm resolving resolution), and two SL4 High Power Tungsten Halogen and Deuterium Lamps (190 - 2500 nm spectral range, 3000 K) were purchased from StellarNet, Inc. The visible portion of the lamps were operated on the third color temperature setting. A 3-way split fiber optic reflection probe was positioned above and normal to the surface of the sample, which was connected to the BLACK-Comet spectrometer, the DWARF-Star spectrometer, and an SL4 lamp (reflection lamp). A collimating lens was positioned below and normal to the surface of the sample, and was connected to the second SL4 lamp (transmission lamp) via a second fiber-optic cable. A mechanical shutter was placed between the collimating lens and the sample, which was darkened with black flocked paper (Thorlabs part number BFP1). The BLACK-Comet UV-Vis and DWARF-STAR Miniature NIR spectrometers were controlled by a Raspberry Pi 3 Model B+ (2017) running Raspbian Stretch (Kernel 4.14) and Python 2.7.0. The SL4 lamps were controlled by an Arduino Due (A000062), which was slaved to the Raspberry Pi.

To perform a transmission measurement the mechanical shutter was opened, the upper reflection lamp internal shutter was closed, and the lower transmission lamp internal shutter was opened. To perform a reflection measurement the mechanical shutter was closed, the lower transmission lamp internal shutter was closed, and the upper reflection lamp internal shutter was opened.

**Spectrometer calibration**

A 75 × 25 × 1 mm glass slide coated with 50 nm aluminum was purchased from Deposition Research Lab Inc. for use as a reflectance baseline. The true specular reflectance of the prepared reference sample was measured with an Agilent Cary 7000 Universal Measurement Spectrometer (UMS) using the Cary Universal Measurement Accessory (UMA) to hold the sample at 10° from normal. The sensitivity of the BLACK-Comet and DWARF-Star spectrometers were set by increasing the integration time (in ms) of the detectors until the signal was between 80% and 95% of saturation, where saturation was $2^{16}$ counts. For transmission measurements, the sensitivity was determined with no sample present, and for reflection measurements, the sensitivity was determined with the calibrated aluminum mirror. The bright and dark baselines for transmission were completed with no sample present and with the transmission lamp on and off, respectively. The bright and dark baselines for reflection were completed with the calibrated aluminum mirror present and with the reflection lamp on and off, respectively. The known true reflection of the aluminum mirror, obtained from the Cary 7000, was used to define the bright reflection baseline.

**Robotic spectroscopy measurement**

For each film fabricated by the robotic platform, UV-Vis-NIR spectra were collected by the robot, which holds samples in the optical path of the spectrometer using the vacuum substrate handler. First the reflection and transmission spectra of a blank glass substrate were collected, followed by analogous reflection and transmission spectra of the annealed thin film on an identical substrate. The spectra of the uncoated and coated substrate were used to compute an approximation to the absorbance of the thin film, as described in the UV-Vis-NIR data processing section. The spectra of each thin film and the spectra of the blank substrate were measured at seven positions spaced ~1 mm apart near the center of the substrate.

**UV-Vis-NIR data processing**

Reflection and transmission spectra were measured at normal incidence and were assumed to be entirely specular and incoherent. This assumption is reasonable as long as surfaces and interfaces scatter a minimal amount of light, and interference fringes in the spectra are minimal. At any wavelength/energy, the raw reflection ($R_0$) and transmission ($T_0$) of the blank substrate can thus be related to the reflectivity/transmissivity ($R_g/T_g$) of the glass-air interface, and to the single-pass transmission of the glass substrate ($X_g$) using the following equations:

$$R_0 = R_g + \frac{R_g X_g^2 T_g^2}{1 - R_g^2 X_g^2}$$

$$T_0 = \frac{X_g T_g^2}{1 - R_g^2 X_g^2}$$

$$1 = R_g + T_g$$

These equations can be solved for $R_g$, $T_g$, and $X_g$.

Since a thin film on a glass substrate is a multilayer system, additional assumptions are needed to process the film/glass spectra analytically. In this work, the refractive indices of the film and substrate were both expected to be ~1.5, and thus the reflection at the film-glass interface could be ignored without significant distortion of the result. This simplification allowed for the raw reflection ($R_1$) and transmission ($T_1$) of the film/substrate to be incorporated into a similar set of equations as above while introducing only three new parameters:

$$R_1 = R_f + \frac{R_g X_g^2 X_f^2 T_f^2}{1 - R_g R_f X_g^2 X_f^2}$$

$$T_1 = \frac{X_g X_f T_g T_f}{1 - R_g R_f X_g^2 X_f^2}$$

$$1 = R_f + T_f$$

In this second set of equations, $R_f$ and $T_f$ are the reflectivity and transmissivity of the film-air interface, respectively, and $X_f$ is the single-pass transmission of the thin film. Solving these equations for $X_f$ gives the corrected transmission of the thin film. The corresponding absorption of the film is:

$$Abs_{film} = -log(X_f)$$

This quantity can be calculated at each measured wavelength/energy to give the corrected absorption spectrum of the film at each of the seven measured positions.

***Autonomous workflow step 6: 4-point probe conductance instrumentation and characterization***

4-point probe conductivity measurements were performed with a Keithley Series K2636B System SourceMeter® Instrument with a Signatone Four Point Probe Head (part number

SP4-40045TBN, 0.040" tip spacing, 45 gram pressure, tungsten carbide tips with 0.010" radii) connected through a Signatone Triax to BNC feedthrough panel (part number TXBA-M160-M).

The current on the outer probes was stepped from 0 to 4 nA in 0.8 nA steps. The system was stabilized at each step for 0.5 s, and the potential across the inner probes was integrated for 25 power line cycles (at 60 Hz). The slope of the potential as a function of the current sourced afforded the resistance. No correction factors were applied to the resistance measurement, as the size of the slide is significantly larger than the spacing between the probes. The conductance of each film was measured at seven positions spaced ~1 mm apart near the center of the substrate. These positions matched those used in spectroscopy measurements.

*Autonomous workflow step 7: pseudomobility calculation*

Conventional theory describes the conductivity ($\sigma$) of p-type semiconductors as the product of the elementary charge (e), the density of positive charge carriers ($\rho_h$), and the mobility of these same charge carriers ($\mu$). In numerous examples, $\rho_h$ is treated as an independent variable programmed by doping fraction, while $\mu$ is an intrinsic material property found to vary with temperature, applied voltage, disorder, and doping fraction (24, 25). $\mu$ thus encompasses most of the complexity of $\sigma$ and is often maximized in order to optimize the performance of electronic materials. In the specific case of a hole transport material (HTM), the doping fraction must be managed carefully given that under-doping reduces hole conductivity, while over-doping risks depletion of valence electrons at the HTM/absorber interface that can lead to inefficient hole injection.

In this work, the value of μ for each HTM film was extracted from sheet resistance 4-point probe and UV-Vis measurements using the following methodology.

The hole mobility of the HTM is expressed as:

$$\mu_h = \frac{\sigma}{\rho_h e}$$

where the hole conductivity of the HTM is assumed to approximate the total conductivity due to the heavy p-doping of most measured conditions and the high intrinsic mobility of holes relative to that of electrons known to exist in spiro-OMeTAD. σ is defined as:

$$\sigma = (R_S \cdot t)^{-1} = \left(\left(\frac{dV}{dI} \cdot C \cdot \pi \cdot t\right)/\ln(2)\right)^{-1}$$

where $R_S$ is the sheet resistance, t is the film thickness, dV/dI is the linear change in voltage (V) with respect to current (I) at low V and I extracted from the 4-point probe measurement, and C is a geometric correction factor tied to the ratio between the probe tip distances and the rectangular dimensions of the film. The redefinition of $R_S$ here is valid for any 4-point probe measurement in which t is significantly less than the distance between adjacent probe tips. $\rho_h$ is defined as:-

$$\rho_h = \rho_{HTM} \cdot \frac{[HTM^+]}{[HTM]}$$

where $\rho_{HTM}$ is the total density of redox active HTM sites in the film and $[HTM^+]/[HTM]$ is the fraction of active sites carrying a positive charge at any given time. This fraction is not necessarily equal to the ratio of dopant to HTM in the film, since not all dopants oxidize HTM material quantitatively. $\rho_{HTM}$ can be converted to a molar concentration by:

$$\rho_{HTM} = 1000 \cdot N_A \cdot [HTM]$$

where $N_A$ is the Avogadro constant and 1000 $N_A$ is the unit of conversion between m$^{-3}$ and M. The ratio of [HTM$^+$] to [HTM] is the same as the ratio of $\rho_h$ to $\rho_{HTM}$. This yields:

$$\rho_h = 1000 \cdot N_A \cdot [HTM^+]$$

which can be used in conjunction with Beer's law to incorporate data from the UV-Vis spectrum of the film in question. The resulting equation:

$$\rho_h = 1000 \cdot N_A \cdot \frac{Abs_{HTM^+}}{\varepsilon \cdot t} = 1000 \cdot N_A \cdot \frac{Abs_{film}}{\varepsilon \cdot t}$$

includes the molar extinction coefficient of the film ($\varepsilon$), the film thickness (t), and the reflection- and substrate-corrected absorbance (Abs$_{HTM+}$) of the film in the wavelength range of 500±5 nm, where all absorption can be attributed to HTM$^+$. This final expression for $\rho_h$ is only valid when the programmed dopant:HTM ratio is at or below 1:1 so that minimal HTM$^{2+}$ exists in the film. Finally, $\mu_h$ can be defined in terms of both experimental results:

$$\mu_h = \frac{\sigma}{\rho_h e} = \left(\frac{ln(2) \cdot \varepsilon}{1000 \cdot \pi \cdot e \cdot C \cdot N_A}\right)\left(\frac{dI}{dV}\right)\left(Abs_{HTM^+}\right)^{-1}$$

This expression is crucially independent of film thickness, allowing for accurate optimization over the full compositional and processing variable space employed in this work. Dividing out the constants in the final equation above yields a parameter termed pseudomobility:

$$pseudomobility = \left(\frac{dI}{dV}\right)\left(Abs_{HTM^+}\right)^{-1}$$

Pseudomobility is equivalent to the quotient of film conductivity and p-type carrier density and provides a useful measure of relative mobility that can be utilized in optimization experiments. The pseudomobility of each thin film was calculated independently at each of the seven characterized positions. The pseudomobility value passed to the optimization algorithm was the mean value of the seven positional pseudomobilities.

*Autonomous workflow step 8: determining the next experiment with Bayesian optimization*

After each thin film experiment, the realized experimental parameters and resulting pseudomobility value were sent to an instance of *ChemOS* (*14*) running on a remote server via a file transfer interface (Dropbox). *ChemOS* then provided the experimental data to the *Phoenics* (*15*) global Bayesian optimization algorithm, initiating an update of the algorithm's surrogate model of the experimental response surface. To minimize the robot's down-time between the completion of one experiment and the initiation of the next experiment, *ChemOS* immediately provides parameters for the next experiment which have been pre-computed by *Phoenics*. In this way, the moderately computationally intensive updating of *Phoenic*'s surrogate model can occur in parallel with the robotic execution of the next experiment, shortening the execution of a campaign with 30 experiments by approximately 30 minutes. *Phoenics* suggests new experiments by using an adjustable sampling parameter to explicitly bias experimental design towards exploration or exploitation in an alternating fashion and thus enables global optimization over the response surfaces explored by the robotic platform. The initial sample is chosen at random as no model of the response surface is initially available. The exchange of information between *ChemOS* and the robotic platform is managed by Python software written in-house.

**Acknowledgments:** The authors would like to acknowledge Maddie Eghtesad for her contributions to the project hardware and experimentation and to Tara Zepel for her contributions in literature research. We thank the UBC Chemistry machine shop for assistance with instrument fabrication. We would like to acknowledge CMC Microsystems for the provision of products and services that facilitated this research, including SolidWorks 2018 SP5.0. For robot control and data processing, we would like to acknowledge the contributors to the Python programming language (Python Software Foundation, https://www.python.org).

**Funding:** The authors thank Natural Resources Canada (EIP2-MAT-001) for their financial support. C. P. B. is grateful to the Canadian Natural Sciences and Engineering Research Council (RGPIN 337345-13), Canadian Foundation for Innovation (229288), Canadian Institute for Advanced Research (BSE-BERL-162173), and Canada Research Chairs for financial support. B. P. M, F. G. L. P., T. D. M., and C. P. B. acknowledge support from the SBQMI's Quantum Electronic Science and Technology Initiative, the Canada First Research Excellence Fund, and the Quantum Materials and Future Technologies Program. J. E. H. is supported by the Canadian Natural Sciences and Engineering Research Council (RGPIN 2016-04613) and Canada Foundation for Innovation (35833). V. L. and H. S. were supported by an NSERC Strategic Partnership Grant (STPGP 493833-16). F. H. acknowledges support from the Herchel Smith Graduate Fellowship and the Jacques-Emile Dubois Student Dissertation Fellowship. L. M. R. and A. A.-G. were supported by the Tata Sons Limited Alliance Agreement (A32391) and the Office of Naval Research (N00014-19-1-2134), and would also like to thank Dr. Anders Frøseth for generous support.

**Author contributions:** A.A-G., J. E. H., and C. P. B. conceived of and supervised the project. B. P. M., M. R., H. S., G. J. N., R. H. Z., and F. G. L. P. developed the robotic hardware. F. G. L. P., T. D. M., J. R. D., T. H. H. and B. P. M. developed the data analysis software. L. P. E. Y., B. P. M., F. G. L. P., G. J. N., R. H. Z., V. L., M. S. E., K. E. D. and H. S. developed the robotic control software. F. H., L. M. R., and A. A.-G. developed the optimization algorithm and interfaced the algorithm with the robot. F. G. L. P., T. D. M., B. P. M., K. E. D., F. H., and L. M. R. designed and performed the robotic optimization experiments. T. D. M, M. S. E., and D. J. D performed additional experiments. All authors participated in the writing of the manuscript.

**Competing interests:** The authors declare no competing interests.

**Data and material availability:** All data needed to evaluate the conclusions in the paper are present in the paper and/or the Supplementary Materials. The raw data recorded by the robotic platform during the two optimization runs is available at https://github.com/berlinguette/ada/tree/master/Science_Advances_aaz8867. Additional data available from authors upon request.

**Supplementary Materials:**

Materials and Methods

Fig. S1. Ada robotic platform for thin film fabrication and characterization.

Fig. S2. Temperature profile of the heating protocol employed by Ada's annealing furnace.

Fig. S3. UV-Vis-NIR film absorptance spectra for both optimization runs.

Fig. S4. Current-voltage relationships for both optimization runs.

Fig. S5. Correlation between pseudomobility measured using Ada and hole mobility measured using hole-only thin film devices.

Fig. S6. Dark field images of highly annealed spiro-OMeTAD thin films with different ratios of dopant.

Fig. S7. Differential scanning calorimetry traces of FK102 Co(III) TFSI salt.

Table S1. Values of manipulated and responding variables for run 1.

Table S2. Values of manipulated and responding variables for run 2.

Table S3. Repeatability of a pseudomobility measurement made using Ada.

Movie S1. Robotic workflow

# Supplementary Materials for

Self-driving laboratory for accelerated discovery of thin-film materials

B. P. MacLeod, F. G. L. Parlane, T. D. Morrissey, F. Häse, L. M. Roch, K. E. Dettelbach, R. Moreira, L. P. E. Yunker, M. B. Rooney, J. R. Deeth, V. Lai, G. J. Ng, H. Situ, R. H. Zhang, M. S. Elliott, T. H. Haley, D. J. Dvorak, A. Aspuru-Guzik*, J. E. Hein*, C. P. Berlinguette*

Correspondence to: cberling@chem.ubc.ca

**This PDF file includes:**

Supplementary materials and methods
Supplementary figures S1 to S7
Supplementary tables S1 to S3
Supplementary movie S1

**Supplementary materials and methods**

*Hole mobility measurement using hole-only devices device fabrication*

Hole-only devices were fabricated and used to perform hole mobility measurements of spiro-OMeTAD films using the steady-state space-charge limited current (SCLC) method[29,30] (see Fig. S5).

Indium tin oxide (ITO)-coated glass substrates (Thin Film Devices, 1.1 mm OLED/OPV grade) were cut into 1.8cm × 3.5 cm pieces. 1.5 cm of the ITO coating was removed by etching with Zn powder and 2M HCl. The etched substrates were then sonicated for 5 minutes in each of the following: detergent (Extran® 300), distilled water, acetone, isopropanol. The ultrasonically cleaned substrates were then blown dry with nitrogen and subjected to a UV-ozone treatment for 30 minutes directly before use.

A PEDOT:PSS ink was formulated by filtering an aqueous dispersion of PEDOT:PSS (Heraeus Clevios™ AI 4083) through a 40 μm PVDF filter and then combining 1 part by volume of this dispersion with 2 parts by volume of isopropanol. PEDOT:PSS films were then manually deposited on the ITO substrates by dynamic spin-coating 100 μL of the PEDOT:PSS ink at 3000 RPM for 1 minute. A Kimwipe ® soaked in water was then used to remove 1 cm of the PEDOT:PSS from each side of the device before baking the device for 1 h at 150 °C.

Spin-coating inks with FK 102 Co(III) TFSI:spiro-OMeTAD molar ratios of 0, 14, 42, 56, 84 and 98% were prepared by robotic pipetting using the *Ada* platform as described in the *Materials and Methods* section of the manuscript. Spiro-OMeTAD films of varying doping levels were then manually deposited on the glass/ITO/PEDOT:PSS devices by dynamic spin coating 50 μL of each ink at 3000 RPM for 1 minute. A Kimwipe soaked in acetone was then used to remove 1 cm of the spiro-OMeTAD film from each side of the device.

To complete the devices, 80 nm-thick gold contact layers were deposited by electron beam evaporation through a Kapton ® shadow mask. The final device area was 3.5 mm$^2$.

Film thicknesses for the HTM layers on the hole-only devices were measured by stylus profilometry (Bruker Dektak XT). Current-voltage curves were obtained for each device using a source-measure unit (Keithley 2400); the applied voltage ranged from 0 to 1 volts and measurements occurred in air. Following Xu et. al[30], , hole mobilities were extracted using the SCLC method by fitting the quadratic region of the current density curve to the Mott-Gurney law

$$J = \tfrac{9}{8}\mu\varepsilon\varepsilon_0 \frac{V^2}{d^3}$$

where $\mu$ is the mobility of the HTM layer, $\varepsilon$ is the relative permittivity of the HTM (for which we assumed a value of 3), $\varepsilon_o$ is the vacuum permittivity and $d$ is the HTM thickness.

With the exception of the device with a dopant:HTM molar ratio of 84% which was removed from the dataset as reliable thickness data was not obtained, the hole mobilities reported in figure S6C are the mean values obtained from all the working devices obtained for each doping level, numbering between 3 - 7. The error bars are the standard deviations across all the measured devices.

*Robotic pseudomobility measurements for comparison to hole mobility measurements*

The pseudomobility values reported in figure S6C are the average of quadruplicate measurements made on HTM films with FK 102 Co(III) TFSI:spiro-OMeTAD molar ratios of 0, 14, 42, 56, 84 and 98%. These films were prepared and characterized using the *Ada* platform as described in the *Materials and Methods* section of the manuscript. The pseudomobility error-bars shown in figure S6C are the standard deviation of the quadruplicate measurement results.

***Determination of the melting point of FK102 Co(III) TFSI by Differential Scanning Calorimetry***

The melting of FK102 Co(III) TFSI salt (see Fig. S7) was measured by a Differential Scanning Calorimeter (Netzsch DSC 214 Polyma) using heating and cooling rates of 10 K min$^{-1}$ and a $N_2$ purge. The sample was heated and cooled five times between 0 °C and 250 °C, with the final two cycles showing a stabilized melting point of 189 °C. The DSC traces from the final two heating cycles are shown in Fig. S7.

# Supplementary figures

All figures containing numerical data were created in Python using the matplotlib library.

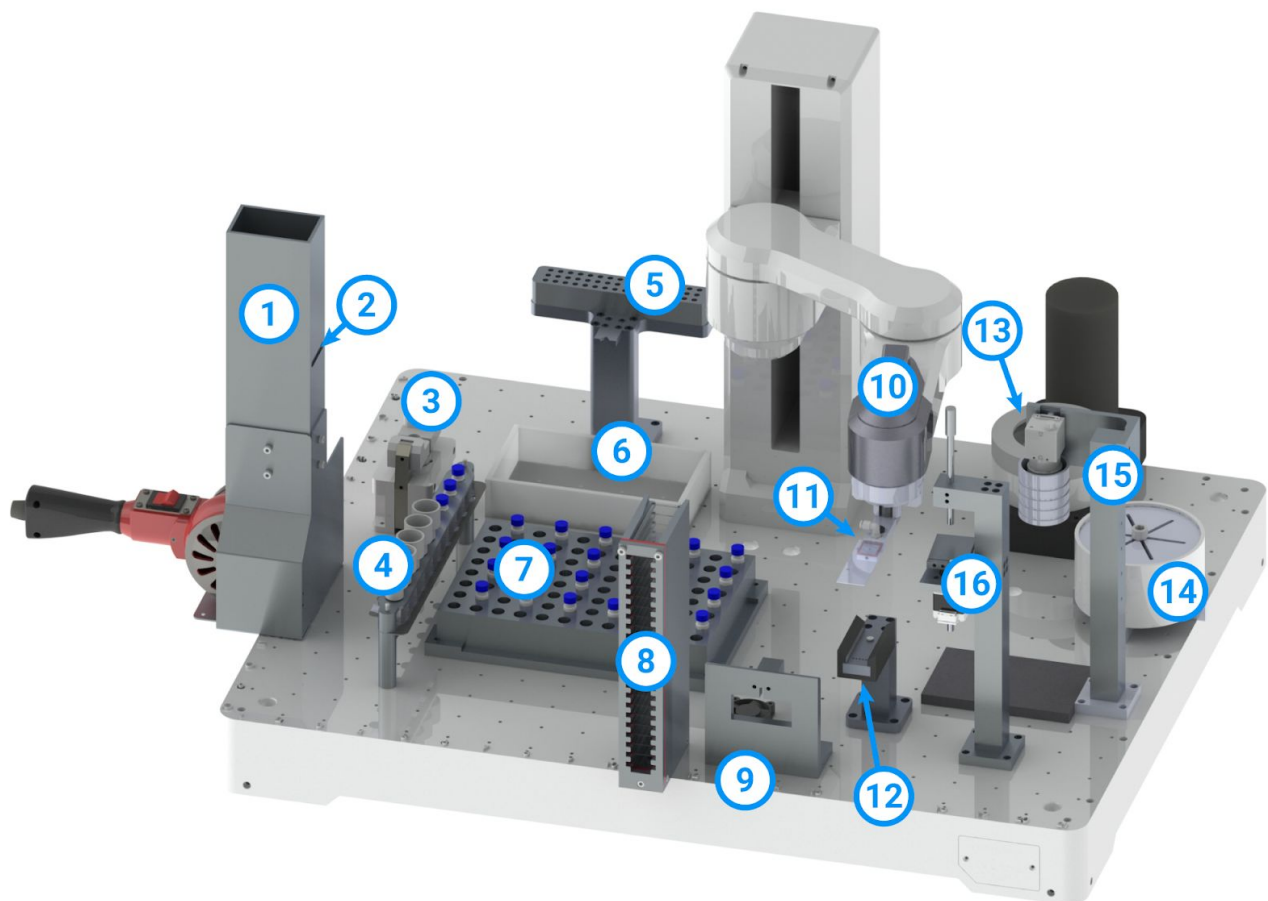

**Fig. S1. *Ada* robotic platform for thin film fabrication and characterization.** This platform includes: (1) an annealing furnace with (2) a slot-shaped sample port; (3) a weigh scale for feedback dispensing of solutions; (4) a rack for stock solution vials; (5) a rack for storing clean pipet tips and (6) a container for disposal of used tips; (7) a rack for clean mixing vials; (8) a rack for clean glass slides; (9) a 4-point probe for measuring film conductance; (10) a robotic arm for handling vials and slides with (11) an attachment for gripping slides and (12) a station for storing this attachment when not in use; (13) a spin coater with (14) a removable lid; (15) a camera for dark field imaging; (16) a spectrometer for transmission and reflection measurements.

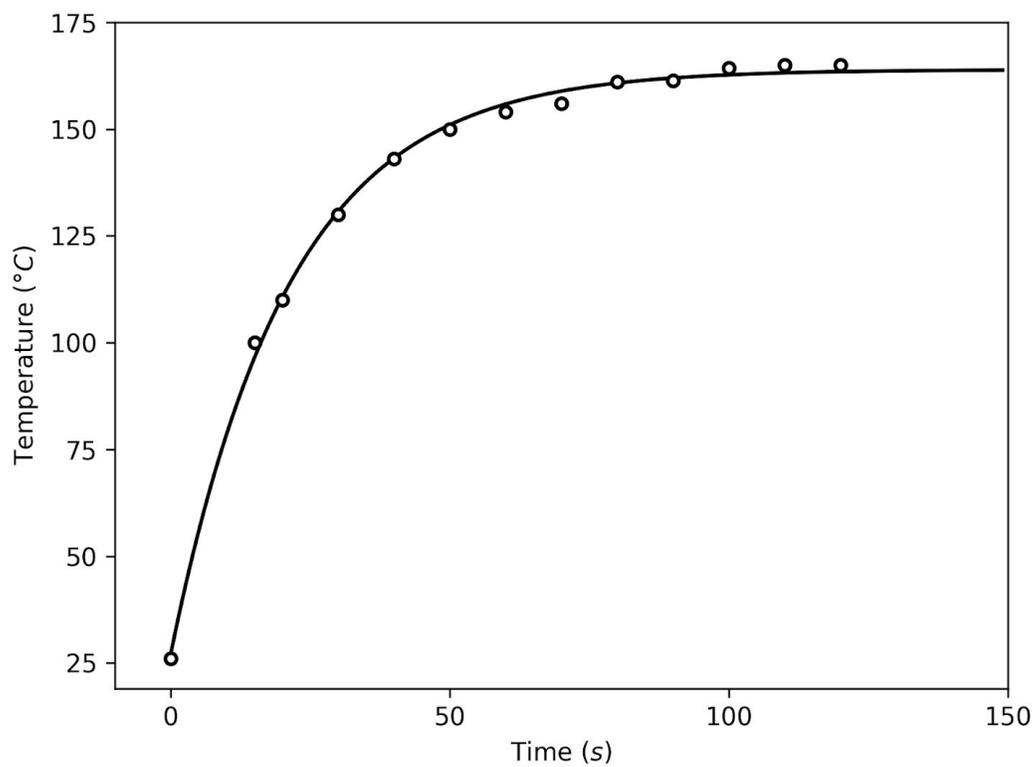

**Fig. S2. Temperature profile of the heating protocol employed by *Ada's* annealing furnace.** A thermocouple was contacted to a glass microscope slide and the measured temperature was collected at a series of times after the heat gun was turned on. The data was fit to an asymptotic regression model.

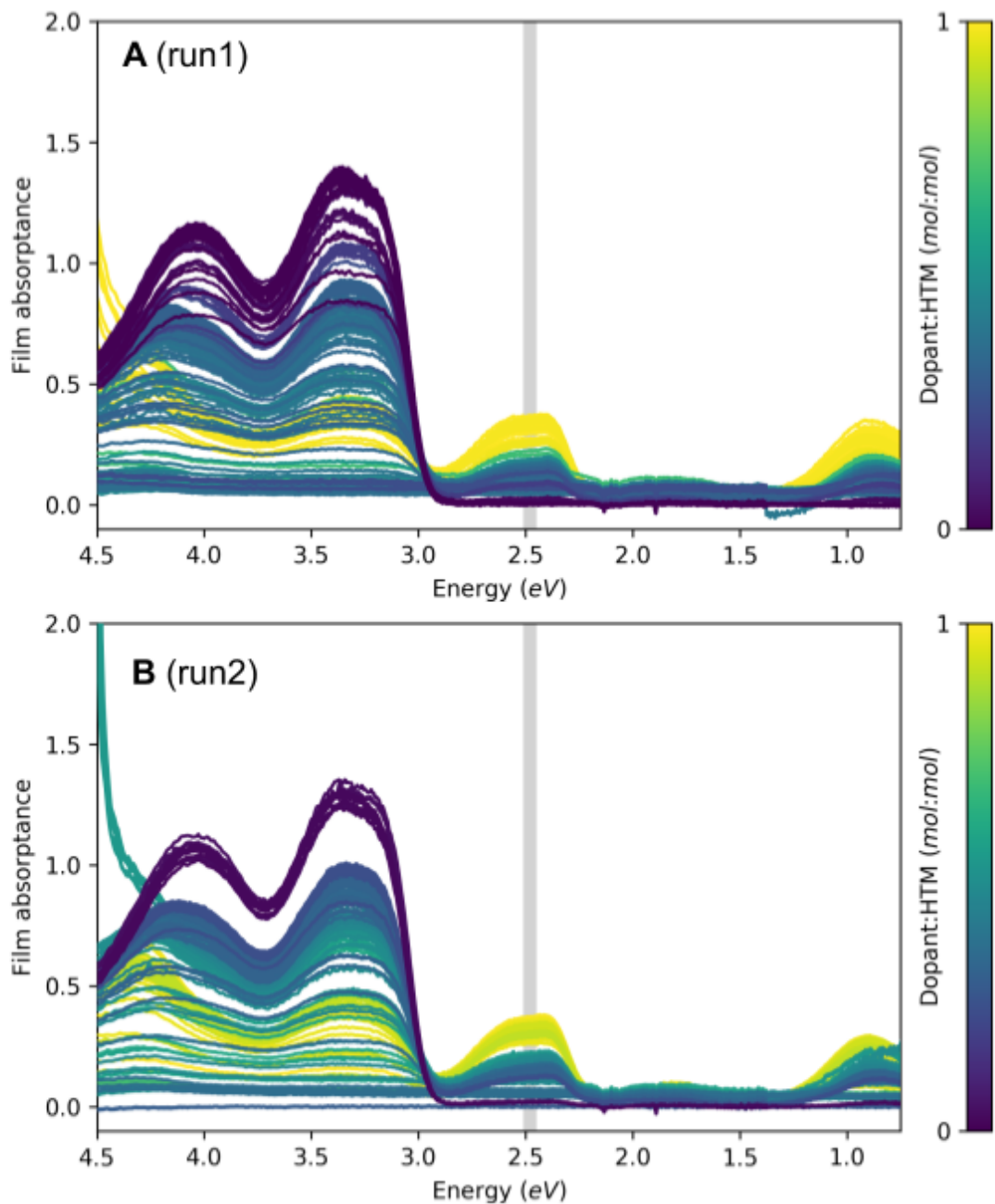

**Fig. S3. UV-Vis-NIR film absorptance spectra for both optimization runs.** (**A**) spectra for run 1. (**B**) spectra for run 2. The film absorption was calculated from the transmission and reflection spectra of both the glass substrate and the deposited film on the glass substrate. Absorption values for films with varying dopant:HTM ratios are shown, as indicated by the side bar. The mean absorption from 495 - 505 nm (indicated by the grey bar) was used to calculate pseudomobility.

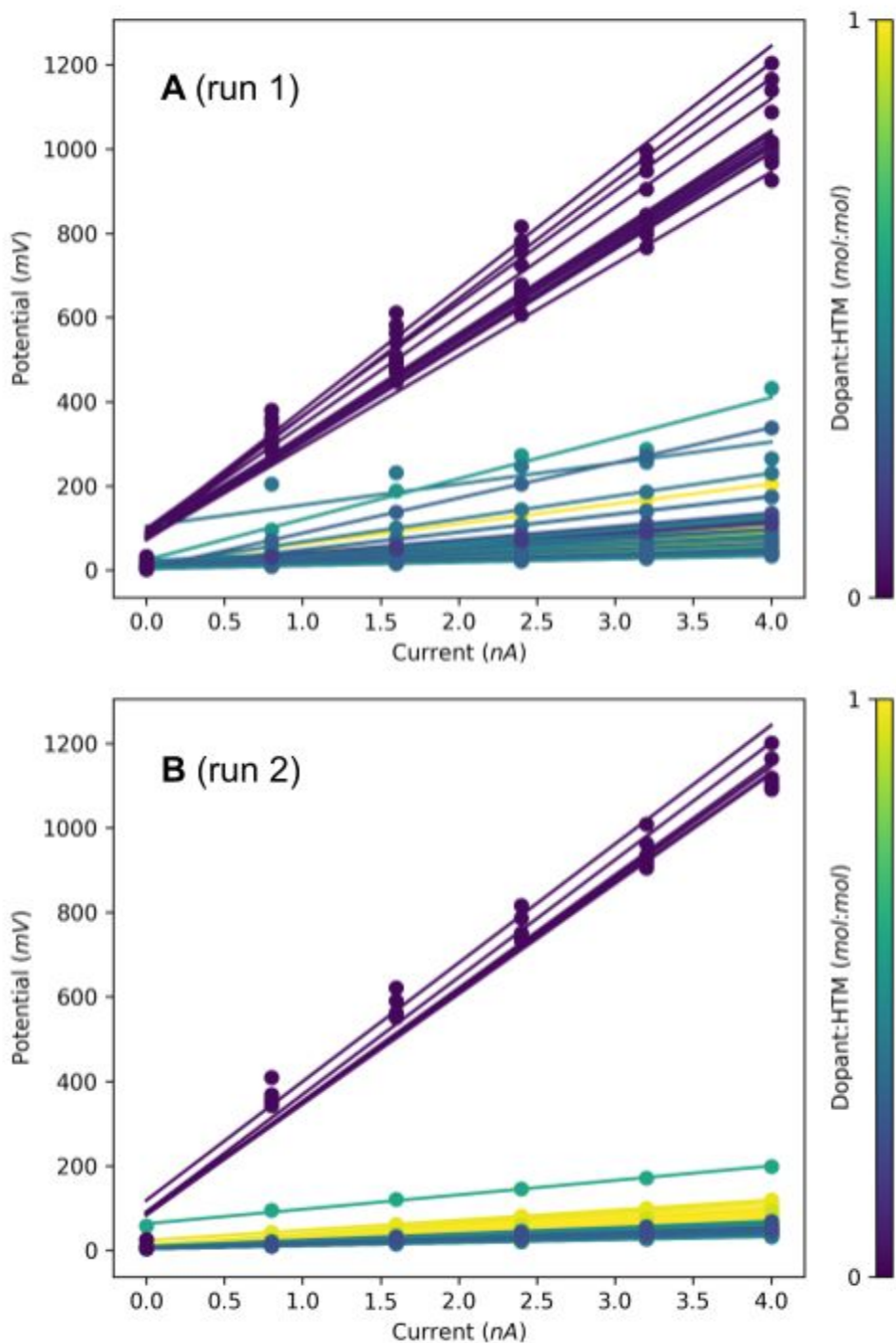

**Fig. S4. Current-voltage relationships for both optimization runs.** (**A**) I-V curves for run 1. (**B**) I-V curves for run 2. Potentials were recorded with a 4-point probe delivering a current between 0 and 4 nA for films with varying dopant:HTM ratios as indicated by the side bar. Conductance was calculated from the fitted slope of the current-voltage plots.

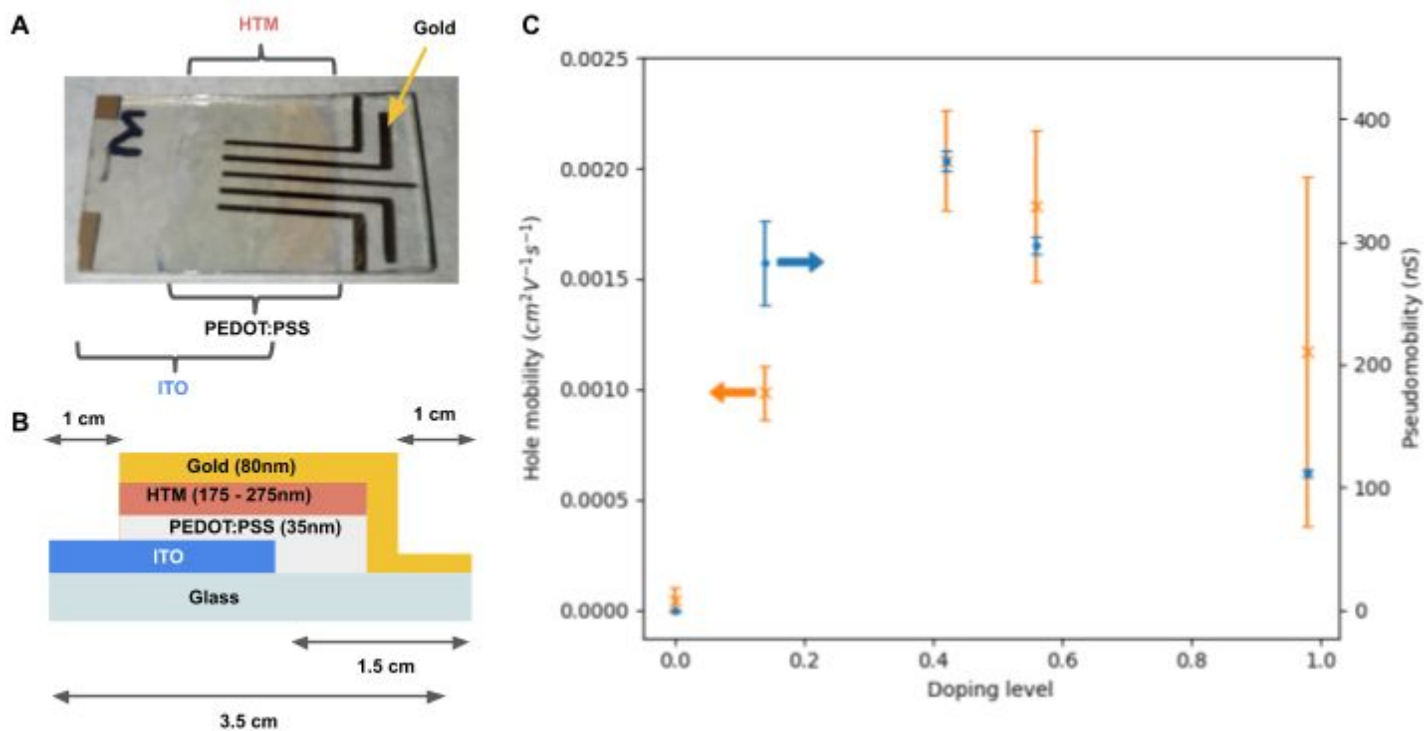

**Fig. S5. Correlation between pseudomobility measured using *Ada* and hole mobility measured using hole-only thin film devices.** (**A**) Photograph of an array of hole-only devices used for measuring hole mobility (**B**) Schematic of the prepared hole-only devices. (**C**) Comparison between manually-measured hole mobility data from the hole-only devices and the pseudomobility data measured using the *Ada* platform. To facilitate visual comparison, the y-axes have been scaled such that the maxima of the two data-sets are equal. Photo credit: David J. Dvorak, The University of British Columbia.

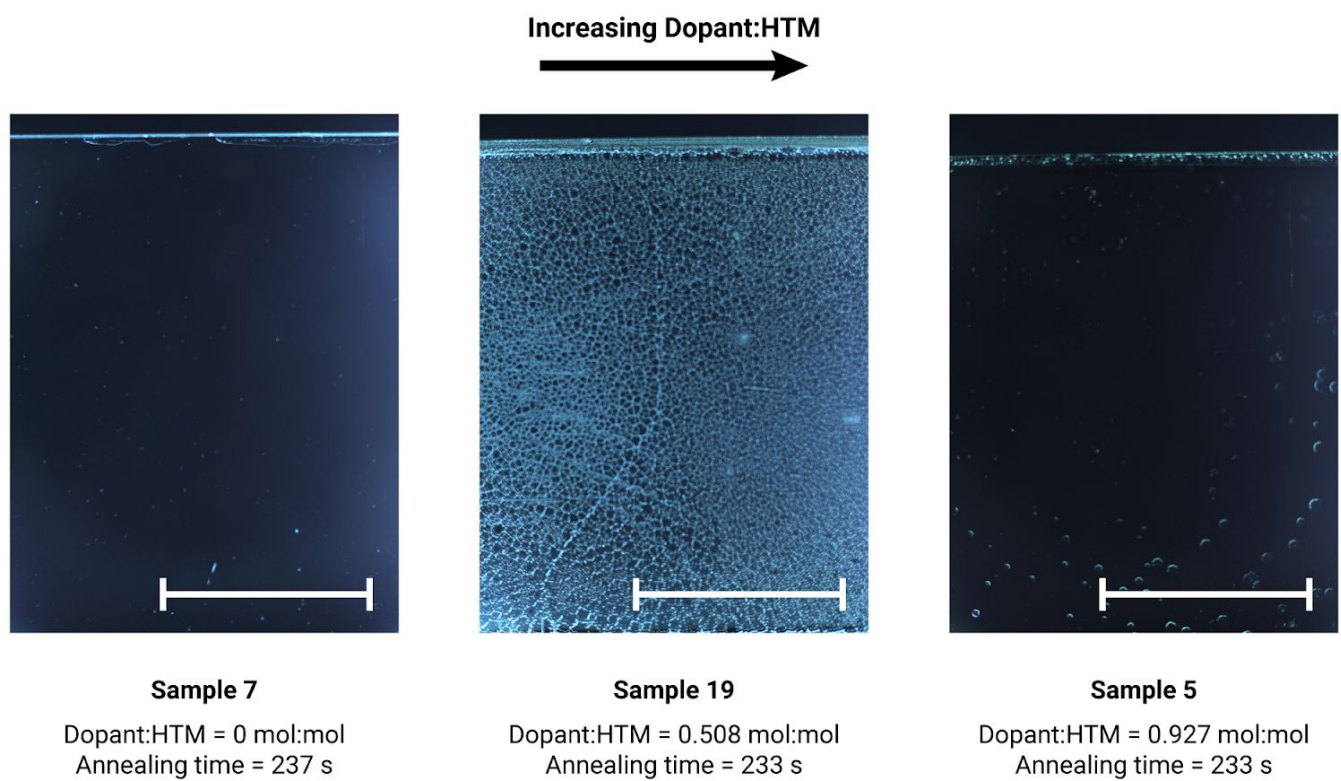

**Fig. S6. Dark field images of highly annealed spiro-OMeTAD thin films with different ratios of dopant.** Shown above are the images of three representative samples from run 2. When no dopant was added, no dewetting was observed. Dewetting was much more significant at an intermediate dopant:HTM ratio compared to a high dopant:HTM ratio. Scale bars are 1 cm.

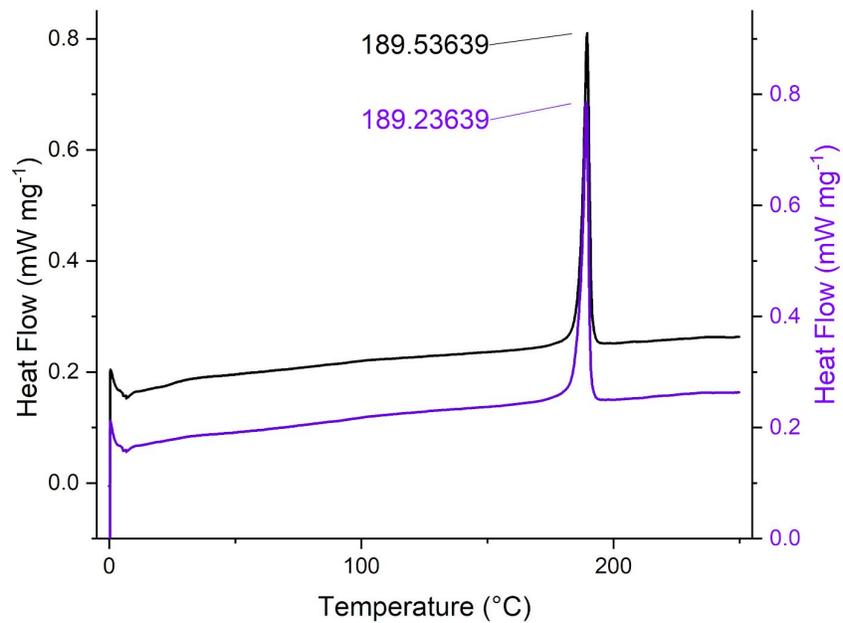

**Fig. S7. Differential scanning calorimetry traces of FK102 Co(III) TFSI salt.** The sharp, labelled endothermic peak in each trace indicates the temperature of the crystalline-to-liquid phase transition. Traces of the fourth and fifth heating cycles are offset vertically for clarity.

**Table S1. Values of manipulated and responding variables for run 1.**

| Sample | Dopant:HTM (mol:mol) | Annealing time (s) | Conductance (nS) | | HTM+ Absorptance at 500 nm | | Pseudomobility (nS) | |
|---|---|---|---|---|---|---|---|---|
| | | | mean | st. dev. | mean | st. dev. | mean | st. dev. |
| 0 | 0.358 | 115 | 7.6 | 20.0 | 0.079 | 0.050 | 47 | 124 |
| 1 | 0.994 | 202 | 50.7 | 3.3 | 0.324 | 0.020 | 157 | 19 |
| 2 | 0 | 237 | 0.0 | 0.0 | 0.001 | 0.000 | 0 | 0 |
| 3 | 0.761 | 133 | 0.0 | 0.0 | 0.067 | 0.004 | 0 | 0 |
| 4 | 0.979 | 1 | 35.9 | 0.3 | 0.258 | 0.006 | 139 | 3 |
| 5 | 0.987 | 227 | 41.8 | 17.9 | 0.314 | 0.038 | 136 | 59 |
| 6 | 0.046 | 2 | 4.2 | 0.1 | 0.026 | 0.000 | 161 | 4 |
| 7 | 0.994 | 216 | 42.2 | 9.6 | 0.329 | 0.033 | 129 | 30 |
| 8 | 0 | 0 | 0.0 | 0.0 | 0.005 | 0.002 | 0 | 0 |
| 9 | 0.178 | 21 | 39.2 | 0.9 | 0.086 | 0.003 | 458 | 23 |
| 10 | 0.491 | 0 | 60.8 | 2.0 | 0.183 | 0.004 | 332 | 11 |
| 11 | 0.341 | 49 | 73.8 | 20.5 | 0.135 | 0.002 | 548 | 156 |
| 12 | 0.394 | 3 | 52.2 | 17.4 | 0.153 | 0.003 | 341 | 114 |
| 13 | 0.444 | 23 | 66.5 | 25.9 | 0.163 | 0.006 | 410 | 165 |
| 14 | 0.025 | 128 | 3.9 | 0.4 | 0.022 | 0.003 | 184 | 25 |
| 15 | 0.402 | 76 | 120.7 | 1.1 | 0.164 | 0.003 | 734 | 13 |
| 16 | 0.458 | 238 | 0.0 | 0.0 | 0.055 | 0.002 | 0 | 0 |
| 17 | 0.409 | 92 | 16.6 | 28.6 | 0.102 | 0.049 | 98 | 168 |
| 18 | 0.522 | 88 | 10.7 | 24.4 | 0.098 | 0.048 | 60 | 129 |
| 19 | 0.395 | 69 | 74.5 | 36.2 | 0.152 | 0.009 | 485 | 237 |
| 20 | 0.389 | 73 | 109.8 | 10.1 | 0.160 | 0.003 | 685 | 52 |
| 21 | 0.387 | 81 | 47.4 | 54.0 | 0.142 | 0.038 | 281 | 319 |
| 22 | 0.975 | 88 | 68.0 | 0.8 | 0.357 | 0.196 | 218 | 59 |
| 23* | 0.376* | 72* | 104.0* | 18.4* | 0.044* | 0.014* | 2,600* | 935* |
| 24 | 0.345 | 73 | 55.9 | 54.8 | 0.153 | 0.008 | 355 | 347 |
| 25 | 0.308 | 82 | 58.1 | 39.3 | 0.139 | 0.006 | 421 | 281 |
| 26 | 0.529 | 172 | 0.0 | 0.0 | 0.073 | 0.010 | 0 | 0 |
| 27 | 0.372 | 70 | 94.9 | 42.9 | 0.163 | 0.003 | 580 | 261 |
| 28 | 0.685 | 75 | 10.4 | 27.6 | 0.129 | 0.064 | 51 | 134 |
| 29 | 0.315 | 63 | 102.4 | 2.5 | 0.141 | 0.004 | 727 | 13 |
| 30 | 0.235 | 167 | 4.5 | 11.8 | 0.090 | 0.030 | 34 | 90 |
| 31 | 0.337 | 72 | 60.8 | 42.0 | 0.138 | 0.021 | 427 | 262 |
| 32 | 0.31 | 81 | 97.7 | 11.1 | 0.141 | 0.008 | 692 | 74 |
| 33 | 0.45 | 58 | 126.6 | 1.5 | 0.180 | 0.003 | 703 | 10 |
| 34 | 0 | 93 | 0.0 | 0.0 | 0.002 | 0.000 | 0 | 0 |

*A brief spectrometer power failure resulted in calibration errors between the spectra of the glass slide and the spectra of the thin film, invalidating the calculation of thin film absorptance. This outlier was removed in all following analyses.

## Table S2. Values of manipulated and responding variables for run 2.

| Sample | Dopant:HTM (mol:mol) | Annealing time (s) | Conductance (nS) | | HTM$^+$ Absorptance at 500 nm | | Pseudomobility (nS) | |
|---|---|---|---|---|---|---|---|---|
| | | | mean | st. dev. | mean | st. dev. | mean | st. dev. |
| 0 | 0.988 | 202 | 53.5 | 3.0 | 0.354 | 0.013 | 152 | 13 |
| 1 | 0.359 | 115 | 9.3 | 24.6 | 0.082 | 0.044 | 58 | 154 |
| 2 | 0.702 | 151 | 0.0 | 0.0 | 0.067 | 0.003 | 0 | 0 |
| 3 | 0.032 | 4 | 2.4 | 0.1 | 0.019 | 0.000 | 122 | 3 |
| 4 | 0.904 | 175 | 77.6 | 6.4 | 0.300 | 0.009 | 259 | 29 |
| 5 | 0.927 | 233 | 46.0 | 28.6 | 0.285 | 0.084 | 190 | 134 |
| 6 | 0.973 | 206 | 39.0 | 21.6 | 0.299 | 0.056 | 139 | 92 |
| 7 | 0 | 237 | 0.0 | 0.0 | 0.008 | 0.016 | 0 | 0 |
| 8 | 0.967 | 212 | 48.9 | 38.7 | 0.284 | 0.102 | 241 | 245 |
| 9 | 0.98 | 1 | 39.3 | 0.1 | 0.275 | 0.002 | 143 | 1 |
| 10 | 1.009 | 147 | 39.8 | 27.1 | 0.290 | 0.068 | 161 | 144 |
| 11 | 0.543 | 0 | 78.5 | 0.5 | 0.204 | 0.003 | 384 | 5 |
| 12 | 1.026 | 159 | 49.4 | 16.9 | 0.318 | 0.016 | 155 | 54 |
| 13 | 0.516 | 1 | 76.9 | 1.2 | 0.192 | 0.005 | 399 | 5 |
| 14 | 0.535 | 43 | 125.6 | 1.0 | 0.219 | 0.005 | 574 | 15 |
| 15 | 0.491 | 0 | 71.0 | 1.7 | 0.189 | 0.008 | 376 | 21 |
| 16 | 0.504 | 58 | 88.2 | 53.4 | 0.190 | 0.007 | 457 | 272 |
| 17 | 0.023 | 127 | 3.7 | 0.1 | 0.023 | 0.001 | 158 | 2 |
| 18 | 0.567 | 67 | 47.3 | 55.9 | 0.170 | 0.051 | 225 | 260 |
| 19 | 0.508 | 233 | 0.0 | 0.0 | 0.055 | 0.004 | 0 | 0 |
| 20 | 0.464 | 47 | 131.5 | 1.3 | 0.185 | 0.003 | 712 | 6 |
| 21 | 0.592 | 59 | 100.6 | 36.7 | 0.191 | 0.035 | 515 | 151 |
| 22 | 0.472 | 41 | 130.2 | 1.6 | 0.184 | 0.002 | 707 | 14 |
| 23 | 0.337 | 70 | 108.5 | 7.4 | 0.155 | 0.009 | 701 | 27 |
| 24 | 0.482 | 62 | 126.7 | 15.4 | 1.556 | 2.310 | 438 | 287 |
| 25 | 0.314 | 70 | 107.3 | 5.6 | 0.149 | 0.008 | 722 | 42 |
| 26 | 0.441 | 50 | 133.3 | 0.4 | 0.182 | 0.002 | 734 | 7 |
| 27 | 0.239 | 79 | 57.5 | 39.9 | 0.122 | 0.001 | 470 | 326 |
| 28 | 0.469 | 66 | 47.4 | 51.2 | 0.154 | 0.039 | 269 | 277 |
| 29* | 0.292* | 57* | 0.0* | 0.0* | 0.001* | 0.000* | 0* | 0* |
| 30 | 0.313 | 28 | 92.7 | 1.3 | 0.138 | 0.003 | 671 | 11 |
| 31 | 0.329 | 58 | 109.0 | 2.7 | 0.146 | 0.003 | 745 | 18 |
| 32 | 0.367 | 19 | 87.1 | 0.9 | 0.148 | 0.002 | 587 | 13 |
| 33 | 0.247 | 59 | 88.9 | 0.6 | 0.119 | 0.001 | 748 | 7 |
| 34 | 0.314 | 34 | 90.5 | 3.7 | 0.127 | 0.007 | 717 | 54 |

*An alignment error during spin coating resulted in no precursor solution deposited onto the glass slide. This outlier was removed in all following analyses.

**Table S3. Repeatability of a pseudomobility measurement made using *Ada*.** The Pseudomobility values shown below were measured on 10 replicate samples with nominal doping ratio 0.247 and annealing time 59 seconds. These samples were robotically prepared and characterized by the Ada platform in a single run over the course of 3.75 hours

| Replicate number | Pseudomobility (nS) |
|---|---|
| 1 | 662.5 |
| 2 | 667.7 |
| 3 | 640.4 |
| 4 | 655.8 |
| 5 | 643.8 |
| 6 | 651.1 |
| 7 | 659.7 |
| 8 | 663.9 |
| 9 | 657.5 |
| 10 | 658.2 |
| *mean* | 656.1 |
| *standard deviation* | 8.687 |
| *standard deviation / mean* | 1.324% |

**Supplementary movies**

Movie S1 - Robotic workflow. For high-resolution version, see: https://youtu.be/0wVLjVdrYEE

This video shows the *Ada* robotic platform performing workflow steps identical or similar to those used to perform the experiments reported in the manuscript.